\documentclass[english]{article}
\usepackage[T1]{fontenc}
\usepackage[latin9]{inputenc}
\usepackage{geometry}
\usepackage[active]{srcltx}
\usepackage{amsmath}
\usepackage{amssymb}
\usepackage{graphicx}

\makeatletter
\newcommand{\lyxaddress}[1]{
\par {\raggedright #1
\vspace{1.4em}
\noindent\par}
}

\@ifundefined{showcaptionsetup}{}{%
 \PassOptionsToPackage{caption=false}{subfig}}
\usepackage{subfig}
\makeatother

\usepackage{babel}
\begin{document}

\title{Direct algebraic mapping transformation for decorated spin models}

\author{Onofre Rojas%
\thanks{correspondig Author email: ors@dex.ufla.br; Phone: +5535 38291954.%
} and S. M. de Souza}

\maketitle

\lyxaddress{Departamento de Ciencias Exatas, Universidade Federal de Lavras,
CP 3037, 37200-000, Lavras - MG, Brazil}
\begin{abstract}
In this article we propose a general transformation for decorated
spin models. The advantage of this transformation is to perform a
direct mapping of a decorated spin model onto another effective spin
thus simplifying algebraic computations by avoiding the proliferation
of unnecessary iterative transformations and parameters that might
otherwise lead to transcendental equations. Direct mapping transformation
is discussed in detail for decorated Ising spin models as well as
for decorated Ising-Heisenberg spin models, with arbitrary coordination
number and with some constrained Hamiltonian's parameter for systems
with coordination number larger than 4 (3) with (without) spin inversion
symmetry respectively. In order to illustrate this transformation
we give several examples of this mapping transformation, where most
of them were not explored yet. 
\end{abstract}

\section{Introduction}

Exactly solvable models is one of the most challenging topics in statistical
physics and mathematical physics. Statistical physics models in general
cannot be solved analytically, but only numerically. For example,
Ising models with spin-1/2 or higher under external magnetic field
are challenging current issues. Exact solutions were obtained only
for a very limited cases. After the Onsager solution for the two-dimensional
Ising model\cite{onsager}, several attempts to solve other similar
models were performed, mainly the honeycomb lattices\cite{Horiguchi-Wu,Kolesik}.
The exact solution for the honeycomb lattice with external magnetic
field was also studied by Wu\cite{Wu-jmp73}, and the Kogome lattice
was also discussed in references \cite{azaria,lu-wu}. Using the method
proposed by Wu\cite{wu-pla86a}, Izmailian\cite{Izmailian-96} obtained
an exact solution for a spin-3/2 square lattice with only nearest
neighbor and two-body spin interactions. Izmailian and Ananikian\cite{Izmailian-96,izm-ank}
have also obtained an exact solution for a honeycomb lattice with
spin-3/2. The Blume-Emery-Griffiths (BEG)\cite{BEG} model for the
honeycomb lattice was also investigated by Horiguchi\cite{Horiguchi-Wu},
Wu \cite{wu-pla86b}, Tucker\cite{tucker} and Urumov\cite{urumov-hnycmb},
following the standard decoration transformation\cite{Fisher,syozi}
and satisfying the Horiguchi condition\cite{Horiguchi-Wu}. The Ising
model on pentagonal lattice was investigated by Waldor et al.\cite{waldor}
and Urumov\cite{urumov}. Some exact results have been obtained with
restricted parameters for spin-1 Ising model by Mi and Yang\cite{MiYang}
using a non-one-to-one transformation\cite{Kolesik}. Some half-odd-integer
spin-$S$ Ising models were already discussed in the literature\cite{Tang}.
Following this line of thought, we have recently found a family of
solutions for half-odd-integer spin\cite{our-pla-09}, where by means
of simple projections we obtain a family of results, a particular
case of which recovers the previous results found in the literature\cite{Izmailian-96,izm-ank,Tang}.

Several decorated Ising models have been solved
using the well-known decoration transformation presented in the 1950's
by M. E. Fisher\cite{Fisher} and Syozi\cite{syozi}, which has been
recently generalized in reference \cite{PhyscA-09} for arbitrary
spin and for any \textit{mechanical} spin, such as the classical-quantum
spin model. This transformation has been widely used in the literature
and, in some cases, it has been applied in several steps that introduce
a number of intermediate parameters such as discussed in reference\cite{urumov,loh,strecka-triang,urumov-hnycmb}.
The decoration transformation can also be applied to classical-quantum
(hybrid) spin models, i.e. Ising-Heisenberg models. Several quasi-one
dimensional models such as the diamond-like chain have been widely
investigate in the literature \cite{strecka-cond-mat09,strecka-conf,strecka06,vadim-10,vadim-nos,vadim-triangl,valvede-diamond},
as well as two-dimensional lattice spin models by using the decoration
transformation approach\cite{our-4-6-latt,strecka-2dim-hybrd,strecka-2dim-quart,strecka-bathroom,strecka-kagome},
which has been successfully applied even to three-dimensional decorated
systems\cite{3dim-decoration}. Another interesting application of
decoration transformation was also investigated in a work by Pereira
et al.\cite{pereira-prb77} in which they considered a delocalized
interstitial electrons on diamond-like chain and also investigated
the magnetocaloric effect in a kinetically frustrated diamond chain\cite{pereira-prb78}.
Meanwhile, Strecka et al.\cite{strecka-2dim-hybrd} discussed the
localized Ising spins and itinerant electrons in two-dimensional models,
as well as two-dimensional spin-electron models with coulomb repulsion\cite{galisova-strck}.
Recently, the decoration transformation approach has been also applied
to spinless interacting particles, thus showing the possibility of
application to interacting electron models\cite{spinless-ferm}. Due
to these important progresses, recently Strecka\cite{strecka-pla}
discussed this transformation in more detail, following the approach
presented in reference \cite{PhyscA-09} for the the case of hybrid
models.

On the  other hand, real materials such as Heterotrimetallic  [DyCuMoCu]$_\infty$ polymer\cite{diana} can be formulated as ising-Heisenberg chain models\cite{heuvel}, as well as Dy$_4$Cr$_4$ complex\cite{rink}  as decorated Ising ring.  This work also was motivated due several synthesized materials, with
more involving complex structure, such as the following materials:
Yb$_{3}$AuGe$_{2}$In$_{3}$: An ordered variant of the YbAuIn structure
exhibiting mixed-valent Yb behavior\cite{comp1}; Density functional
theory analysis of the interplay between Jahn-Teller instability,
uniaxial magnetism, spin arrangement, metal-metal interaction, and
spin-orbit coupling in Ca$_{3}$CoMO$_{6}$ (M = Co, Rh, Ir)\cite{comp1a};
supramolecular Co(II)-{[}2$\times$2{]} grids: metamagnetic behavior
in a single molecule\cite{comp3}; magnetic ordering in Iron tricyanomethanide\cite{comp4};
spin frustration in MII{[}C(CN)$_{3}${]}$_{2}$ (M = V, Cr). magnetism
and neutron diffraction study\cite{comp5}.

In this paper we present a direct generalized transformation for a
mixed or decorated spin model onto a uniform spin model, in which
the main difference to the aforementioned generalized transformation\cite{Fisher,PhyscA-09,strecka-pla}
is that there is no step by step transformation. The seminal idea
of decorated spin model transformation of type star-star already was
emphasized and used in a particular case by Baxter\cite{baxter-86},
as well as by M. E. Fisher\cite{Fisher66jmp} to study the planar
Ising model using dimer solution. In order to introduce a direct transformation
of decorated spin model onto a uniform spin-1/2 model, we will follow
the basic idea used by Baxter\cite{baxter-86}. In order to illustrate
this transformation we consider the decoration transformation displayed
in figure \ref{fig:2-leg-trans}:

\begin{figure}[h]

\centering{}\includegraphics[scale=0.4]{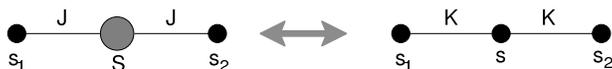}\caption{\label{fig:2-leg-trans} The decoration transformation of the mixed
spin-(1/2,$S$) Ising model onto its equivalent uniform spin-1/2 Ising
model.}
\end{figure}

The Boltzmann factor for decorated spin model or mixed spin-(1/2,$S$)
model (see figure \ref{fig:2-leg-trans}) could be expressed by
\begin{equation}
w(s_{1},s_{2})=\sum_{\mu=-S}^{S}\mathrm{e}^{J(s_{1}+s_{2})\mu},\label{eq:w12-two}
\end{equation}
 in which $S$ is any spin value larger than $1/2$, whereas $s_{1}$and
$s_{2}$ are the spin values of the spin-1/2 particles. For simplicity,
$J$ represents here the spin-spin coupling in units of $-\beta=-1/kT$,
in which $k$ is Boltzmann's constant while the $T$ means the absolute
temperature. From now on we will use this convenient notation for
all parameters of the Hamiltonian.

The Boltzmann factor for the effective uniform spin-1/2 model, as
displayed in figure \ref{fig:2-leg-trans}, could be expressed by
\begin{equation}
\tilde{w}(s_{1},s_{2})=\sum_{s=\pm\frac{1}{2}}\mathrm{e}^{K_{0}+K(s_{1}+s_{2})s},\label{eq:w-eff-2-leg}
\end{equation}
 where $K$ represents the spin-spin interaction parameter in units
of $-\beta$, whereas $K_{0}$ is a \textquotedbl{}constant\textquotedbl{}
shift energy in units of $-\beta$. The term $\mathrm{e}^{K_{0}}$
also could be understood as the Z-invariant factor\cite{baxter-86,Fisher66jmp}.
In eqs. \eqref{eq:w12-two} and \eqref{eq:w-eff-2-leg} we assume
that the spin-inversion symmetry is satisfied, i.e., the system remains
invariant under reversion of all spins.

In order that both spin models become equivalent, we impose the following
relation $\tilde{w}(s_{1},s_{2})=w(s_{1},s_{2})$, for the Boltzmann
factors of the effective spin model and of the decorated spin model.
For the spin-1/2 case we obtain two equations assuming the spin-inversion
symmetry is satisfied. We obtain 
\begin{equation}
2\mathrm{e}^{K_{0}}\cosh\left(K/2\right)=\sum_{i=-S}^{S}\cosh\left(iJ\right),\label{eq:trns-1}
\end{equation}
 for the configuration $\uparrow\uparrow$, whereas for the configuration
$\uparrow\downarrow$ we have
\begin{equation}
2\mathrm{e}^{K_{0}}=2S+1,\label{eq:trns-2}
\end{equation}
 Therefore the constant $K_{0}$ is obtained easily from eq. \eqref{eq:trns-2},
whereas the parameter $K$ can be obtained from eqs. \eqref{eq:trns-1}
and \eqref{eq:trns-2}.

The present work aims at showing that this transformation could be
easily extended for any $q$-leg decorated or mixed spin models, $q\in\{3,4,\dots\}$,
mapping it onto a uniform $q$-leg star spin-1/2 Ising models. It
is organized as follows. In sec. 2 we present the generalized $q$-leg
star-star transformation. In sec. 3 we discuss this transformation
without spin-inversion symmetry. In sec. 4 it is extended to higher
values of spin. In sec. 5 we discuss the transformation for quantum-classical
spin and in sec. 6 we present our conclusions.

\section{The generalized $q$-leg star-star decoration transformation}

\begin{figure}[h]
 \centering{}\includegraphics[scale=0.4]{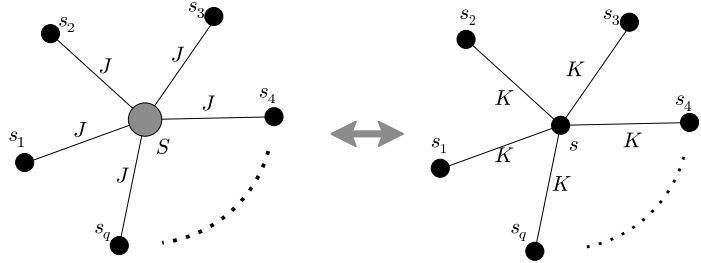}\caption{\label{fig:q-leg-trans}The q-leg star-star transformation of the
mixed spin-(1/2,$S$) Ising model onto its equivalent uniform spin-1/2
Ising model.}
\end{figure}

Following the transformation proposed in the previous section, we
can perform a transformation assuming a general coupling for $q$-leg
star spin model as illustrated in figure \ref{fig:q-leg-trans}, under
total spin-inversion invariance. The Hamiltonian for $q$-leg star
spin model, in units of $-\beta$ may be expressed as

\begin{equation}
H=\sum_{j=1}^{[S]}\left(J_{2j-1}\mu^{2j-1}\sum_{i=1}^{q}s_{i}+D_{2j}\mu^{2j}\right),\label{eq:Ham-gen}
\end{equation}
 where $\left[S\right]$ means the largest integer less than or equal
to $S$, by $J_{j}$ we mean the coupling coefficient of $\mu^{j}s_{i}$,
whereas $D_{j}$ is the coupling coefficient of $\mu^{j}$ with $\mu=\{-S,...,S\}$.
Note that eq.\eqref{eq:Ham-gen} is invariant under total spin-inversion
($\sum_{i=1}^{q}s_{i}\rightarrow-\sum_{i=1}^{q}s_{i}$) and $\mu\rightarrow-\mu$.

Therefore the decorated $q$-leg star spin model Boltzmann factor
could be written as
\begin{equation}
w\left(\left\{ s_{i}\right\} \right)=\sum_{\mu=-S}^{S}\exp\left(\sum_{j=1}^{[S]}\left(J_{2j-1}\mu^{2j-1}\sum_{i=1}^{q}s_{i}+D_{2j}\mu^{2j}\right)\right),\label{eq:wn-dec}
\end{equation}
 in which by $\{s_{i}\}$ we mean the set of $\{s_{1},s_{2},...,s_{q}\}$.

On the other hand, we conveniently consider the star spin-1/2 Ising
model, since models involving spin-1/2 systems could be transformed
onto exactly solvable models\cite{baxter}. Therefore the Boltzmann
factor for uniform spin-1/2 Ising model with zero magnetic field is
given by
\begin{equation}
\tilde{w}\left(\left\{ s_{i}\right\} \right)=\sum_{s=\pm\frac{1}{2}}\exp\left(K_{0}+K(\sum_{i=1}^{q}s_{i})s\right).\label{eq:w-eff}
\end{equation}
 The spin legs interacting with the central spin only depends on $\varsigma\equiv\sum_{i=1}^{q}s_{i}$,
then the Boltzmann factor \eqref{eq:wn-dec} is rewritten as
\begin{equation}
w(\varsigma)=\sum_{\mu=-S}^{S}\exp\left(\sum_{j=1}^{[S]}\left(J_{2j-1}\mu^{2j-1}\varsigma+D_{2j}\mu^{2j}\right)\right),\label{eq:w-dec-gen}
\end{equation}
 and their respective Boltzmann factor \eqref{eq:w-eff} in the effective
spin model becomes
\begin{equation}
\tilde{w}\left(\varsigma\right)=\sum_{s=\pm\frac{1}{2}}\mathrm{e}^{K_{0}+K\varsigma s}.\label{eq:w-eff-gen}
\end{equation}

Therefore the Boltzmann factors are conveniently and simply denoted
by $w(\varsigma)$ and $\tilde{w}(\varsigma)$. For higher spin this
notation is not any more valid, so, in that case we will consider
explicitly each spin contribution.

In general, we can rewrite the result presented in the introduction
as a function of Boltzmann factors of the decorated spin model, imposing
the equivalence of both Boltzmann factors $\tilde{w}(s_{1},s_{2})=w(s_{1},s_{2})$,
which in its turn yields 
\begin{align}
2\mathrm{e}^{K_{0}}= & w(0),\label{eq:w(0)}\\
2\mathrm{e}^{K_{0}}\cosh\left(K/2\right)= & w(1).\label{eq:w(1)}
\end{align}

Note that $w(0)$ and $w(1)$ must be obtained from eq. \eqref{eq:w-dec-gen}.

Then the solution of such algebraic system equation is expressed by
\begin{align}
K_{0}= & \ln\left(\frac{w(0)}{2}\right),\label{eq:rK0}\\
K= & 2\ln\left(\tfrac{w(1)}{w(0)}\pm\sqrt{\left(\tfrac{w(1)}{w(0)}\right)^{2}-1}\right).\label{eq:rK}
\end{align}

This transformation is equivalent to a double decoration transformation\cite{Fisher,PhyscA-09}.
The advantage of the direct mapping in this case is to avoid the unnecessary
intermediate parameters (see references \cite{Fisher,PhyscA-09})
introduced that would only make the calculation more cumbersome\cite{urumov,loh}
and, in some cases leading to an apparently transcendental equation.

\subsection{Three-leg star-star transformation}

In order to solve equations \eqref{eq:w-dec-gen} and \eqref{eq:w-eff-gen}
for the case of three-leg star spin model, we need to assume the spin-inversion
symmetry; so, we have only two configurations $\uparrow\uparrow\uparrow$
and $\uparrow\uparrow\downarrow$, which correspond to $\varsigma=3/2$
and $\varsigma=1/2$, respectively. Once we have two algebraic equations
with two unknown parameters $K_{0}$ and $K$, then we are able to
solve the algebraic system equations, therefore the transformation
can be performed exactly for arbitrary parameter values of decorated
spin models.

The unknown parameters in the effective uniform spin-1/2 model will
be expressed in terms of all arbitrary parameters of the decorated
spin model, assuming that both models are equivalent which means $\tilde{w}(\varsigma)=w(\varsigma)$,
thus, we have, 
\begin{align}
2\mathrm{e}^{K_{0}}\cosh\left(K/4\right)= & w(1/2),\\
2\mathrm{e}^{K_{0}}\cosh\left(3K/4\right)= & w(3/2),
\end{align}
 thus, the solution of the algebraic system equations can be written
explicitly by 
\begin{align}
K= & 2\ln\left(\tfrac{w(3/2)}{w(1/2)}+1\pm\sqrt{\left(\tfrac{w(3/2)}{w(1/2)}+1\right)^{2}-4}\right)-2\ln(2),\label{eq:3-legs-K}\\
K_{0}= & \frac{1}{2}\ln\left(\frac{w(1/2)^{3}}{3w(1/2)+w(3/2)}\right).\label{eq:3-legs-K0}
\end{align}

Once again this transformation is equivalent to a double transformation
(something like as star-triangle-star transformation). By using a
star-star direct transformation\cite{baxter-86}, we avoid the introduction
of unnecessary intermediate parameters (such as the intermediate parameters
to represent the triangle structure system) customary in the literature
(e.g. \cite{urumov,loh}) making the mapping more easy to manipulate.

\subsection{Four-leg star-star transformation}

Another important transformation is the four-leg decorated spin model,
in which there are three spin configurations for the legs: $\uparrow\uparrow\uparrow\uparrow$,
$\uparrow\uparrow\uparrow\downarrow$ and $\uparrow\uparrow\downarrow\downarrow$.
Under total spin-inversion symmetry, any permutations and inversions
of spin always fall into one of these three configurations. Using
the notation $\varsigma=s_{1}+s_{2}+s_{3}+s_{4}$, these three configuration
correspond just to $\varsigma=0$, 1 and 2, respectively. Assuming
both Boltzmann factors are equivalent $\tilde{w}(\varsigma)=w(\varsigma)$,
the algebraic systems equation becomes,
\begin{align}
2\mathrm{e}^{K_{0}}= & w(0),\label{eq:w4(0)}\\
2\mathrm{e}^{K_{0}}\cosh\left(K/2\right)= & w(1),\label{eq:w4(1)}\\
2\mathrm{e}^{K_{0}}\cosh\left(K\right)= & w(2).\label{eq:w4(2)}
\end{align}

The first two equations eqs. \eqref{eq:w4(0)} and \eqref{eq:w4(1)}
are identical to those found for the case of two-leg transformation
which is given by eq. \eqref{eq:w(0)} and \eqref{eq:w(1)} respectively.
For the case of four-leg star-star transformation, we have one additional
equation given by \eqref{eq:w4(2)} , but similarly to the previous
case, there are only two unknown parameters. In order to satisfy completely
the algebraic system of equations we need to impose the following
additional relation between Boltzmann factors of the decorated spin
model, yielded by the manipulation of eqs. (\ref{eq:w4(0)}-\ref{eq:w4(2)}),
\begin{equation}
w(0)w(2)+w(0)^{2}=2w(1)^{2}.\label{eq:eqv-8-vrtx}
\end{equation}
 This means that, in general, at most two parameters of a decorated
spin model could be constrained.

\begin{figure}[h]

\begin{centering}
\includegraphics[scale=0.4]{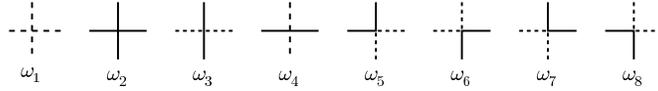} 
\par\end{centering}

\raggedright{}\caption{\label{fig:8-vertex}A line representation of the eight-vertex\cite{baxter}
model formulated for the four-leg star.}
\end{figure}

It is interesting to highlight that eq.\eqref{eq:eqv-8-vrtx} was
obtained only using algebraic manipulation in order to satisfy the
algebraic system of equations given by \eqref{eq:w4(0)}-\eqref{eq:w4(2)}.
Surprisingly the relation \eqref{eq:eqv-8-vrtx} represent nothing
but the special case of the \textit{free fermion condition} of 8-vertex
model\cite{baxter} formulated for four-leg star. The 8-vertex model
configuration displayed in figure \ref{fig:8-vertex} can be compared
with eq.\eqref{eq:eqv-8-vrtx} by the following relations $\omega_{1}=w(2)$,
$\omega_{2}=\omega_{3}=\omega_{4}=w(0)$ and $\omega_{5}=\omega_{6}=\omega_{7}=\omega_{8}=w(1)$
.

In principle, some mixed spin with four-leg star and constrained parameters
can be mapped onto an exactly solvable rectangular Ising model \cite{baxter}.
In what follows we discuss this mapping for some particular case.

For integer values of the central spin, such as spin-1, eq.\eqref{eq:eqv-8-vrtx}
lead us only to a trivial solution ($J_{1}=0$). For spin-2, the Boltzmann
weights are obtained from eq. \eqref{eq:w-dec-gen}, then eq. \eqref{eq:eqv-8-vrtx}
reduce to

\begin{equation}
2t^{16}\left(y^{3}x-1\right)^{2}\left(xy^{7}-1\right)^{2}\left(y^{6}x^{2}+y^{3}x+1\right)^{2}r^{4}+t^{15}\left(xy^{4}-1\right)^{4}\left(xy^{4}+1\right)^{4}r^{3}+x^{2}y^{14}\left(xy-1\right)^{4}=0,\label{eq:sol-spn-2-1}
\end{equation}
 where $x=\exp(J_{1})$, $y=\exp(J_{3})$, $r=\exp(D_{2})$ and $t=\exp(D_{4})$.
Thus we can verify that eq.\eqref{eq:sol-spn-2-1} becomes a quartic
equation in relation to the variable $r$, and their coefficients
are all non-negatively defined, therefore once again we only obtain
a trivial solution for $J_{1}=0$ and $J_{3}=0$. We expect this property
should predominate for higher order integer spins, in accordance with
those discussed in reference \cite{our-pla-09}, where the integer
spin cannot be mapped onto a spin-1/2 Ising model.

However for half-odd-integer central spin we obtain some non-trivial
solutions. 
\begin{itemize}
\item For spin-3/2 the condition \eqref{eq:eqv-8-vrtx} becomes
\begin{equation}
r^{10}\left(y^{7}x-1\right)^{2}\left(xy^{13}+1\right)^{2}\left(xy^{13}-1\right)^{2}\left(y^{7}x+1\right)^{2}\left(x^{2}y^{14}+1\right)^{2}=0,\label{eq:sol-spin-3/2}
\end{equation}
 with $x=\exp(J_{1}/2)$, $y=\exp(J_{3}/8)$ and $r=\exp(D_{2}/4)$.
From eq.\eqref{eq:sol-spin-3/2} we obtain a non-trivial results,
recovering the previous results obtained by a different approach,
for more detail see reference \cite{Izmailian-96,our-pla-09}.
\item For spin-5/2 the eq. \eqref{eq:eqv-8-vrtx}, may be expressed in a
similar way to the previous cases. After some tedious algebraic manipulation
we have
\begin{align}
0= & r^{3}t^{39}\left(x^{4}y^{76}z^{1684}-1\right)^{2}\left(xy^{49}z^{1441}-1\right)^{2}+r^{2}t^{34}\left(x^{2}y^{62}z^{1562}-1\right)^{2}\left(x^{3}y^{63}z^{1563}-1\right)^{2}\nonumber \\
 & +x^{2}y^{98}z^{2882}\left(x^{2}y^{14}z^{122}-1\right)^{2}\left(xy^{13}z^{121}-1\right)^{2},\label{eq:spin-5/2}
\end{align}
 where $x=\exp(J_{1}/2)$, $y=\exp(J_{3}/8)$, $z=\exp(J_{5}/32)$,
$r=\exp(D_{2}/4)$ and $t=\exp(D_{4}/16)$. The eq. \eqref{eq:spin-5/2}
is a cubic equation in relation to the variable $r$ and all coefficients
of eq.\eqref{eq:spin-5/2} are non-negative, therefore there is no
positive solution for $r$, unless all coefficients becomes simultaneously
zero. In order to find the solutions of eq. \eqref{eq:spin-5/2} we
chose the following possibility, let us assume that $x^{4}y^{76}z^{1441}=1$
from the coefficient of term $r^{3}$ and $x^{2}y^{62}z^{15632}=1$
from the coefficients of $r^{2}$, thus we have $y=z^{-30}$ and $x=y^{-31}z^{-781}=z^{149}$.
Satisfying this conditions the eq. \eqref{eq:spin-5/2} is identically
zero. This expression in terms of the Hamiltonian's parameter becomes
as $J_{3}=-\frac{15}{2}J_{5}$ and $J_{1}=\frac{149}{32}J_{5}$, thus
the Boltzmann factor given by eq. \eqref{eq:w-dec-gen} reduce to
\begin{equation}
w(\varsigma)=\sum_{\mu=-5/2}^{5/2}\exp\left(J_{5}\left(\tfrac{149}{16}\mu-\tfrac{15}{2}\mu^{3}+\mu^{5}\right)\varsigma\right)=\sum_{\mu=-5/2}^{5/2}\mathrm{e}^{J_{5}\frac{5!}{16}\sigma(\mu)\varsigma},
\end{equation}
where $\sigma(\mu)$ is defined by 
\begin{equation}
\sigma(\mu)=\frac{149}{120}\mu-\mu^{3}+\frac{2}{15}\mu^{5},
\end{equation}
with $\mu$ we represent the Ising spin-5/2. Note that $\sigma(\mu)$
only takes special values $\{\frac{1}{2},-\frac{1}{2},\frac{1}{2},-\frac{1}{2},\frac{1}{2},-\frac{1}{2}\}$
when $\mu$ takes $\{5/2,3/2,1/2,-1/2,-3/2,-5/2\}$ respectively.
Using similar process for other combination of constrained parameters,
we obtain additional solutions from eq. \eqref{eq:spin-5/2}, recovering
all the three solutions obtained previously in reference \cite{our-pla-09}
using a different mapping (for detail see the eq. (25) of reference
\cite{our-pla-09}). 
\end{itemize}
Following the present mapping, we could recover the mapping for higher
half-odd-integer spins obtained previously in reference \cite{our-pla-09},
where was used a different approach, furthermore we should probably
obtain even additional results to those already found in \cite{our-pla-09},
as was shown here for spin-5/2.

\subsection{General condition for $q$-leg star-star transformation}

For general $q$-leg Ising spin star-star transformations with central
spin $S$ it is possible to obtain the solution for arbitrary $q$.
\begin{itemize}
\item In order to satisfy the condition of star-star transformation, first
we consider the case of even values of $q$ ($q\geqslant4$). We have
two unknown parameters and $q/2$ algebraic equations. Therefore we
must have $(q/2-2)$ conditions to be satisfied for the decorated
$q$-leg star spin model, which read 
\begin{align}
2w(\tfrac{r}{4})^{2}=w(0)\left(w(\tfrac{r}{2})+w(0)\right); & \text{ for }\tfrac{r}{2}:\;\mathrm{even},\label{eq:ws-cond-even-e}\\
2w(\tfrac{r}{4}-\tfrac{1}{2})w(\tfrac{r}{4}+\tfrac{1}{2})=w(0)\left(w(\tfrac{r}{2})+w(1)\right); & \text{ for }\tfrac{r}{2}:\;\mathrm{odd},\label{eq:ws-cond-even-o}
\end{align}
 with $r=\{4,6,8,\ldots,q\}$. It is worth to notice that this condition
depends on the even or odd character of $r/2$.
\item On the other hand, for the case of $q$ odd (for $q\geqslant5$),
we still have two unknown parameters and $([q/2]-2)$ algebraic system
equations, so the parameters of the original $q$-leg star Ising spin
model must satisfy the following conditions,
\begin{equation}
w(\tfrac{r}{2}-1)\left(w(\tfrac{r}{2}-1)+w(\tfrac{r}{2}-3)\right)=w(\tfrac{r}{2}-2)\left(w(\tfrac{r}{2})+w(\tfrac{r}{2}-2)\right),\label{eq:ws-cond-odd}
\end{equation}
 where $r=\{5,7,9,\ldots,q\}$. For the case of $q$ odd we only have
one kind of relation for a given odd $r$.
\end{itemize}
As we can see, the number of constrained parameters increases with
the number of legs or the coordination number, whereas the maximum
number of coupling parameter increases with the spin $S$. The condition
for any arbitrary $q$-leg spin are identical to those of the ($q-2$)-leg
decoration transformation plus one additional condition that only
appears when $q$-leg spin is considered. In other words, all conditions
on ($q-2$)-leg are valid for $q$-leg star plus one new additional
condition that involves $w(\tfrac{q}{2})$ as displayed by eqs.\eqref{eq:ws-cond-even-e},
\eqref{eq:ws-cond-even-o} and \eqref{eq:ws-cond-odd}.

This transformation should correspond to the double star-polygon-star
transformation proposed in reference \cite{PhyscA-09}, where the
polygons involve long-range interactions. However, the decoration
transformation proposed in \cite{Fisher,syozi,PhyscA-09} leads to
a cumbersome coupling, whereas the direct transformation proposed
here just needs to satisfy the conditions \eqref{eq:ws-cond-even-e}
and \eqref{eq:ws-cond-even-o} for $q$ even, and the condition \eqref{eq:ws-cond-odd}
must be satisfied for $q$ odd.

\section{Star-star transformation without spin-inversion symmetry}

The transformation previously discussed could be easily extended for
the $q$-leg star spin model without spin-inversion symmetry too.
For a decorated or mixed spin model transformed onto a uniform spin-1/2
$q$-leg star Ising model with external magnetic field, its decorated
$q$-leg star Boltzmann factor can be written in a similar way as
in eq. \eqref{eq:wn-dec}, in which the spin legs interacting with
decorated spin depend only on $(\varsigma=\sum_{i=1}^{q}s_{i})$ which
we can denote for simplicity only by $\varsigma$, thus yielding
\begin{equation}
w(\varsigma)=\sum_{\mu=-S}^{S}\exp\left(\sum_{j=1}^{2S}\left(J_{j}\mu^{j}\varsigma+D_{j}\mu^{j}-B\varsigma/q\right)\right),\label{eq:w-dec-gen-no-inv}
\end{equation}
with $J_{j}$ and $D_{j}$ are coupling parameters, and $B$ represents
the external magnetic field on the legs. The effective Boltzmann factor
for an uniform star Ising spin model is expressed by
\begin{equation}
\tilde{w}\left(\varsigma\right)=\sum_{s=\pm\frac{1}{2}}\mathrm{e}^{K_{0}+K\varsigma s-h\varsigma/q-h_{0}s}.\label{eq:w-eff-gen-no-inv}
\end{equation}
where $K_{0}$ is a constant energy, $K$ is the coupling term, with
$h$ and $h_{0}$ being the external magnetic field for the legs and
central spin respectively.

A particular case of this transformation is the two-leg ($q=2$) decoration
transformation without spin-inversion and with $h_{0}=h$ in eq. \eqref{eq:w-eff-gen-no-inv}.
Therefore assuming the Boltzmann factor \eqref{eq:w-dec-gen-no-inv}
is equivalent to \eqref{eq:w-eff-gen-no-inv}, we have $\tilde{w}(\varsigma)=w(\varsigma)$.
For this case we have three equations and three unknown parameters
that must satisfy the following relations

\begin{align}
a= & \frac{w_{0}c}{1+c^{2}},\\
b= & \frac{c\left(w_{-1}-\sqrt{w_{1}w_{-1}-w_{0}^{2}}\right)}{w_{0}},\\
c= & \frac{w_{0}^{2}\pm\delta\sqrt{-w_{0}^{2}+w_{1}w_{-1}}}{w_{0}^{2}+w_{-1}\delta},
\end{align}

in which, for simplicity, the Boltzmann factor is denoted by $w_{\varsigma}\equiv w(\varsigma)$,
and $\delta=w_{-1}-w_{1}$. The variables are defined as $a=\exp(K_{0})$,
$b=\exp(K/2)$ and $c=\exp(-h/2)$.

\begin{figure}[h]

\begin{centering}
\includegraphics[clip,scale=0.3]{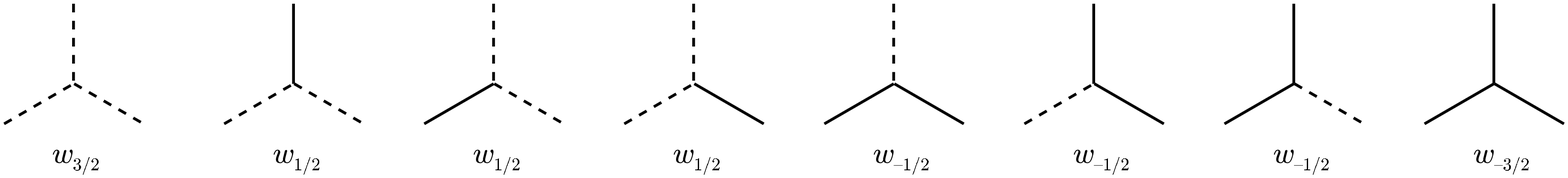} 
\par\end{centering}

\caption{\label{fig:8-vrtx-hnycmb}A line representation of 8-vertex configuration
for three-leg star.}
\end{figure}

Another transformation that we consider is the three-leg star-star
transformation. For this case we have four equations and four unknown
parameters. This transformation is related to the 8-vertex model on
the honeycomb lattice such as discussed by Lin and Wu\cite{Lin-wu1990}.
The schematic representation of 8-vertex model for the honeycomb lattice
is given in figure \ref{fig:8-vrtx-hnycmb}.

Therefore in a similar way to that of the honeycomb lattice we may
express the solutions in terms of the Boltzmann factors as follows,

\begin{align}
\mathrm{e}^{-\tfrac{2}{3}h}= & \frac{w_{3/2}w_{-1/2}-w_{1/2}^{2}}{w_{-3/2}w_{1/2}-w_{-1/2}^{2}},\\
\cosh^{2}(K/2)=\frac{1}{4} & \tfrac{\left(w_{3/2}w_{-3/2}-w_{1/2}w_{-1/2}\right)^{2}}{\left(w_{3/2}w_{-1/2}-w_{1/2}^{2}\right)\left(w_{-3/2}w_{1/2}-w_{-1/2}^{2}\right)},\\
\sinh(h_{0})= & \tfrac{\sinh(K/2)\left(w_{3/2}w_{-1/2}^{3}-w_{-3/2}w_{1/2}^{3}\right)}{\left(w_{3/2}w_{-1/2}-w_{1/2}^{2}\right)\left(w_{-3/2}w_{1/2}-w_{-1/2}^{2}\right)},\\
\mathrm{e}^{4K_{0}}= & \tfrac{\left(w_{3/2}w_{-1/2}-w_{1/2}^{2}\right)\left(w_{-3/2}w_{1/2}-w_{-1/2}^{2}\right)}{16\sinh^{4}(K/2)}.
\end{align}

As we can verify, the effective spin model parameters always may be
expressed in terms of Boltzmann factors. A particular case of this
transformation could be the mixed spin-(1/2,$S$) Ising model on the
honeycomb lattice, which can be transformed onto the spin-1/2 Ising
model with external magnetic field on honeycomb lattice, such as considered
by Azaria-Giacomini\cite{azaria} and Wu\cite{Wu-jmp73,wu-pla86a,wu-pla86b},
where this model is related to 8-vertex model as displayed in fig.\ref{fig:8-vrtx-hnycmb}.

If we consider a particular case of the decorated spin model with
uniform external magnetic field on the honeycomb lattice, i.e. $h_{0}=h$.
This leads to constrained parameters, the constraints of which could
be written as in previous cases, even though more involved.

Finally the extension for coordination number larger than 3 can be
performed straightforwardly, although the conditions of Boltzmann
factors will become more involved expressions.

\section{Star-star transformation for higher spin}

Another interesting case that worth to comment is when the $q$-leg
star-star transformation has spin larger than spin-1/2. Certainly,
this kind of model can be extended in a similar way as in \cite{PhyscA-09};
however, the decoration transformation is subject to more conditions
(and thus more constrained parameters appears) for its validity.

The Hamiltonian for the decorated spin model with $q$-legs could
have a similar treatment to that of sec.2, thus we may write

\begin{equation}
H=\sum_{i=1}^{2S}\sum_{r=1}^{q}\left(\sum_{j=1}^{S_{0}}J_{i,j}\mu_{r}^{i}\mu^{j}-\frac{1}{q}D_{i}\mu_{r}^{i}\right)-\sum_{i=1}^{2S_{0}}B_{i}\mu^{i},\label{eq:Ham-S-orig}
\end{equation}
 in which we use the notation $J_{i,j}$ for the parameter of the
Hamiltonian, which corresponds to the coupling coefficients of the
term $\mu_{r}^{i}\mu^{j}$, whereas by $D_{i}$ we represent the coupling
coefficient of term $\mu_{r}^{i}$, and the last term $B_{i}$ means
the coupling coefficient of the term $\mu^{i}$, with $\mu=\{-S,...,S\}$.

Following the same process developed in sec.2 and illustrated in fig.\ref{fig:q-leg-trans},
the Hamiltonian of the intermediate mixed spin model becomes

\begin{equation}
H^{\prime}=\sum_{i=1}^{2S}\sum_{r=1}^{q}\left(M_{i}\mu_{r}^{i}\sigma-\frac{1}{q}D_{i}^{\prime}\mu_{r}^{i}\right)-h\sigma+M_{0},\label{eq:Ham-S-eff}
\end{equation}
 in which $M_{i}$ represents the parameter of $\mu_{r}^{2i}\sigma$,
whereas $D_{i}^{\prime}$ are the coefficients of the term $\mu_{r}^{i}$,
with $h$ being the external magnetic field strength and $M_{0}$
corresponds to constant energy terms. Using the direct decoration
transformation we can map the system with Hamiltonian \eqref{eq:Ham-S-orig}
onto another effective system with Hamiltonian \eqref{eq:Ham-S-eff}.

It is worth to notice that the Boltzmann factor of higher order spins
does not depend only on $\varsigma=\sum_{i=1}^{q}\mu_{i}$, but also
depends on each spin $\mu_{i}$; consequently, we need to express
explicitly the Boltzmann factor in terms of each spin $\mu_{i}$.

\begin{figure}[h]
 \centering{}\includegraphics[scale=0.3]{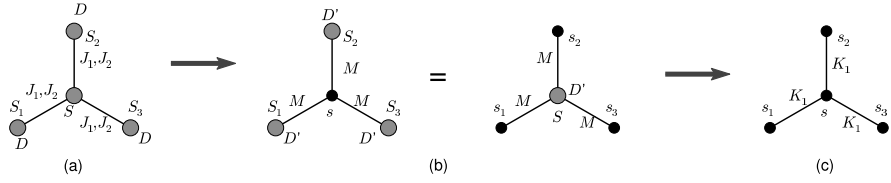}\caption{\label{fig:spin-1-start}Uniform spin-1 star transformation onto its
uniform spin-1/2 star}
\end{figure}

As an illustrative example let us consider the uniform spin-1 ($S=S_{0}=1$)
with $q=3$, as displayed in fig.\ref{fig:spin-1-start}(a), transforming
onto the effective model described by left side of fig.\ref{fig:spin-1-start}(b)
($S=1$, $\sigma=\pm1/2$). Under spin inversion symmetry the Boltzmann
factor of decorated spin model becomes,

\begin{equation}
W(\mu_{1},\mu_{2},\mu_{3})=\sum_{\mu=-1}^{1}\exp\left(J_{11}(\mu_{1}+\mu_{2}+\mu_{3})\mu+J_{22}(\mu_{1}^{2}+\mu_{2}^{2}+\mu_{3}^{2})S^{2}-D_{2}(\mu_{1}^{2}+\mu_{2}^{2}+\mu_{3}^{2})/3-B_{1}\mu^{2}\right),\label{eq:W(Ss)}
\end{equation}
 whereas the Boltzmann factor for effective spin models is given by
\begin{equation}
W^{\prime}(\mu_{1},\mu_{2},\mu_{3})=\exp\left(M_{0}+-D_{2}^{\prime}(\mu_{1}^{2}+\mu_{2}^{2}+\mu_{3}^{2})/3\right)\sum_{s=\pm\tfrac{1}{2}}\mathrm{e}^{M_{1}(\mu_{1}+\mu_{2}+\mu_{3})s}.
\end{equation}

Using the Boltzmann factor of this elementary cell, we are able to
reproduce i.e. the honeycomb lattice Ising model with spin-1 and mixed
spin-(1,1/2) as illustrated in fig.\ref{fig:spin-1-start}(a) and
fig.\ref{fig:spin-1-start}(b) respectively. Imposing the relation
$W^{\prime}(\mu_{1},\mu_{2},\mu_{3})=W(\mu_{1},\mu_{2},\mu_{3})$,
we have six configurations and three unknown parameters to be determined
for the effective spin-(1,1/2) star. In analogy to the previous case
we must have three identities that must be satisfied by the Boltzmann
factors.

\begin{align}
W(1,1,1)W(0,0,0)= & W(0,1,0)W(1,0,-1),\label{eq:vinc-1}\\
2W(1,1,0)W(1,0,0)= & (W(1,1,1)+W(1,1,-1))W(0,0,0),\\
2W(1,0,0)^{2}= & (W(1,1,0)+W(1,-1,0))W(0,0,0),.\label{eq:vinc-3}
\end{align}

For the Hamiltonian (\ref{eq:Ham-S-orig}) the previous eqs. (\ref{eq:vinc-1}-\ref{eq:vinc-3})
are all equivalent, leading just to one relation for the decoration
transformation the parameters of Boltzmann factor which satisfy the
following relation

\begin{equation}
\exp(J_{22})=\cosh(J_{11}),\label{eq:Hor-cond}
\end{equation}
 which is also known as Horiguchi's condition\cite{Horiguchi-Wu},
obtained using the standard decoration transformation\cite{Fisher,syozi}.

For the dual lattice in fig.\ref{fig:spin-1-start}(b) (right side),
the Boltzmann factor is given by 
\begin{equation}
\tilde{w}^{\prime}(s_{1}+s_{2}+s_{3})=\mathrm{e}^{M_{0}}\sum_{\mu=-1}^{1}\exp\left(M_{1}(s_{1}+s_{2}+s_{3})\mu-D^{\prime}\mu^{2}\right),\label{eq:w' S}
\end{equation}
 and it can be expressed in terms of \eqref{eq:W(Ss)} as

\begin{align}
\tilde{w}^{\prime}(3/2)= & \tfrac{1}{2}W(0,0,0)+W(1,1,1),\\
\tilde{w}^{\prime}(1/2)= & \tfrac{1}{2}W(0,0,0)+W(1,1,-1).
\end{align}

The Hamiltonian of 3-leg star effective spin model could be expressed
as

\begin{equation}
\tilde{H}=K_{0}+K(s_{1}+s_{2}+s_{3})s.
\end{equation}

The decoration transformation will be applied in a similar way to
the previous transformation discussed in section 2 (for details see
fig. \ref{fig:q-leg-trans}), where the spins on the legs have same
values for both models. Then the Boltzmann factor is given by \eqref{eq:w-eff-gen}.

Performing a further direct transformation $\tilde{w}^{\prime}(\varsigma)=\tilde{w}(\varsigma)$,
as illustrated in fig. \ref{fig:spin-1-start}, we obtain the results
given by eqs. \eqref{eq:3-legs-K} and \eqref{eq:3-legs-K0} for $K$
and $K_{0}$ respectively.

Here we showed how the direct transformation could be applied in just
two steps only (see fig. \ref{fig:spin-1-start}), rather than in
five steps via the standard decoration transformation\cite{urumov-hnycmb},
we verify that constrained parameter given by eq. (2) of reference\cite{urumov-hnycmb}
is identical to our eq.\eqref{eq:Hor-cond}.

\section{Decoration transformation for classical-quantum spin models}

The transformation presented in section 2 also can be extended for
classical-quantum (hybrid) spin models such as Ising-Heisenberg models,
following a similar approach proposed recently by Strecka\cite{strecka-pla}.
Here we show how this transformation can be used for a particular
kind of lattice without loosing its general properties.

\subsection{Hybrid-star decoration transformation}

\begin{figure}[h]
\subfloat[]{\includegraphics[scale=0.4]{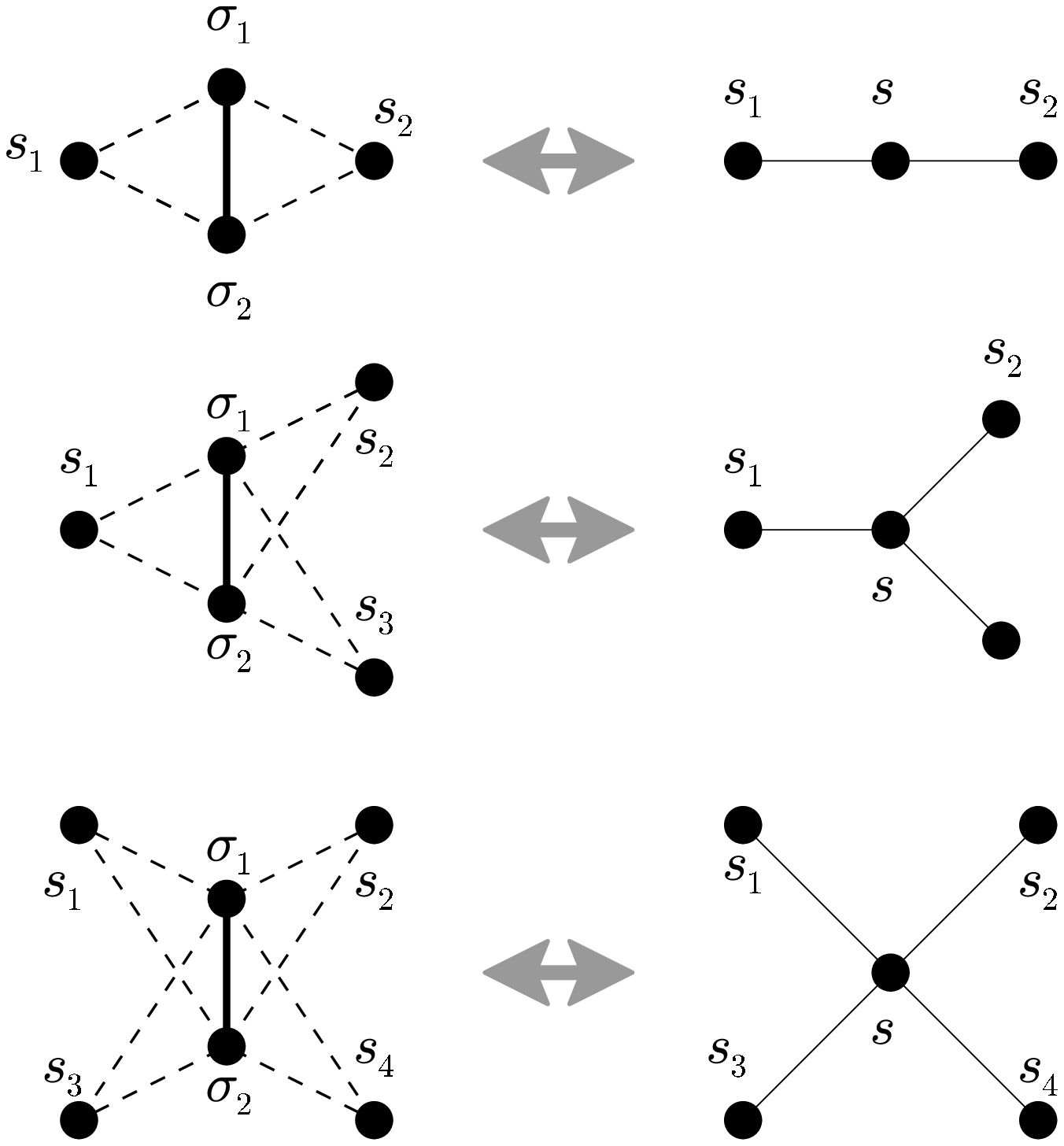}}\hskip1.5cm\subfloat[]{\includegraphics[scale=0.4]{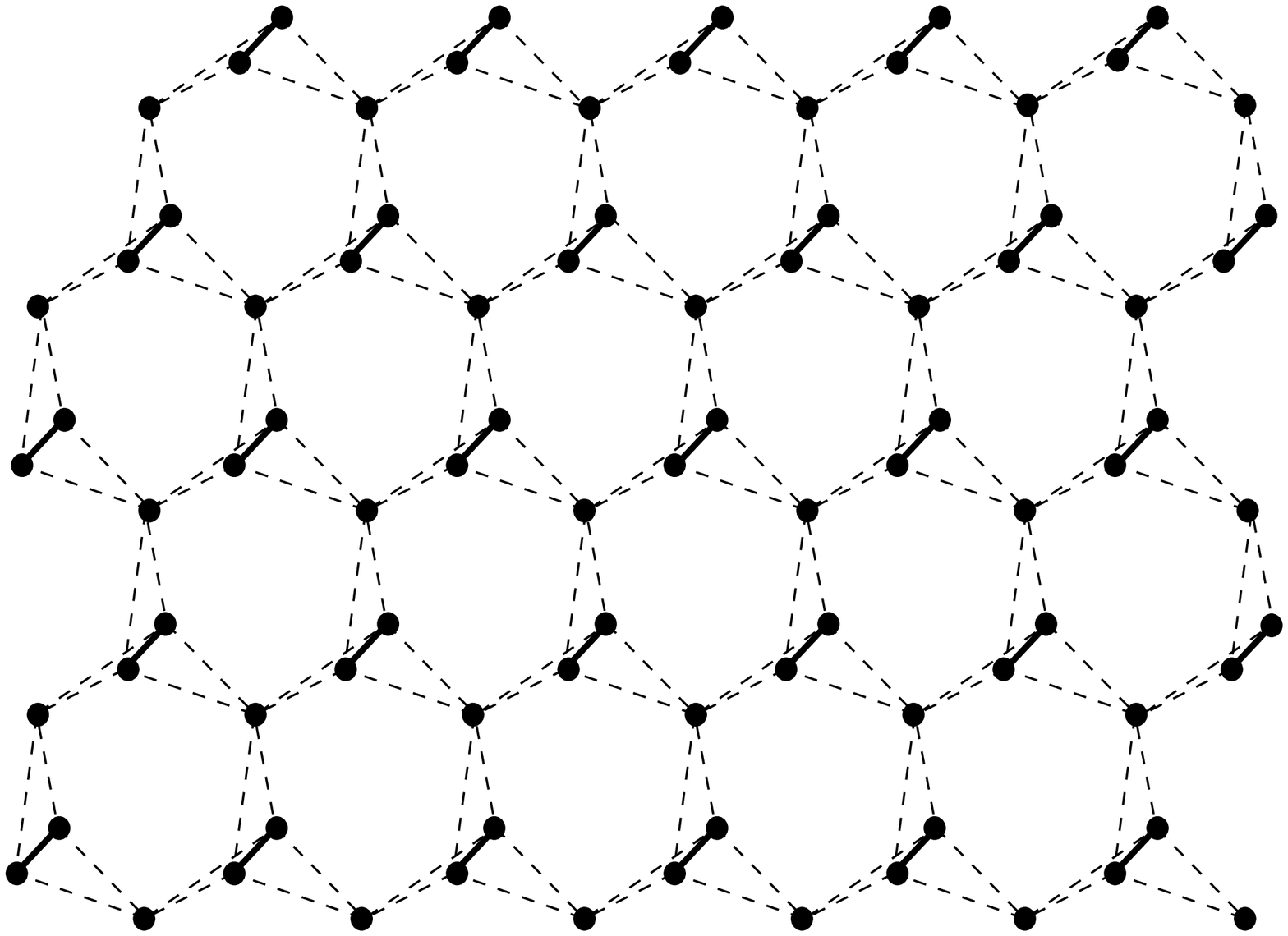}}

\caption{\label{fig:2-leg-hybd}(a) Some examples of hybrid-star decoration
transformation. The thick solid line represents a Heisenberg-like
interaction whereas the dashed line and thin solid line represents
the Ising interaction. In (b) we display a schematic representation
of a lattice, assembled by a second transformation illustrated in
(a).}
\end{figure}

As a first case let us consider the hybrid-star decoration transformation
displayed in figure \ref{fig:2-leg-hybd}(a), in which solid thick
lines represent a Heisenberg-like interaction whereas dashed and solid
thin lines represent the Ising interaction. Thus the Boltzmann factor
may be expressed by
\begin{equation}
w(\{s\})=\mathrm{tr}_{\{\sigma\}}\exp\left(H^{XXZ}(\sigma_{1},\sigma_{2})+J_{2}(\sigma_{1}^{z}+\sigma_{2}^{z})(s_{1}+s_{2}+...+s_{q})\right),\label{eq:hyb-dec}
\end{equation}
 where $H^{XXZ}(\sigma_{1},\sigma_{2})=J_{1}(\sigma_{1}^{x}\sigma_{2}^{x}+\sigma_{1}^{y}\sigma_{2}^{y})+J_{z}\sigma_{1}^{z}\sigma_{2}^{z}$,
and $J_{1}$, $J_{2}$ and $J_{z}$ are interacting parameters of
the Hamiltonian.

In order to calculate the trace in eq. \eqref{eq:hyb-dec} first we
diagonalize the Hamiltonian \eqref{eq:hyb-dec}, and then we obtain
the Boltzmann factor which reads
\begin{equation}
w(\varsigma)=2\mathrm{e}^{J_{z}/4}\cosh(J_{2}\varsigma)+2\mathrm{e}^{-J_{z}/4}\cosh(J_{1}/2).\label{eq:w-hybrd-eff}
\end{equation}

However the effective Boltzmann factors is still expressed by eq.
\eqref{eq:w-eff-gen}. This transformation can be applied to several
types of lattice, mainly in one and two dimensional models. The quasi-two-dimensional
Ising-Heisenberg model represented in fig. \ref{fig:2-leg-hybd}(b)
can be transformed onto an exactly solvable two-dimensional Ising
model using the second transformation illustrated in fig. \ref{fig:2-leg-hybd}(a).

\begin{figure}[h]

\begin{centering}
\includegraphics[scale=0.3]{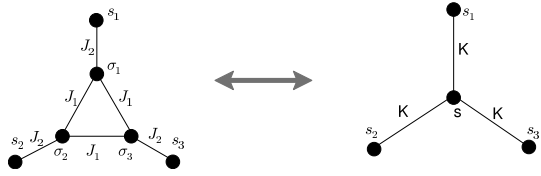} 
\par\end{centering}

\caption{\label{fig:3-leg-hybrid-star}3-leg hybrid-star decoration transformation
onto uniform spin model.}
\end{figure}

Now let us consider another example, a three-leg hybrid-star transformation,
where the decoration consists of three spins forming a triangle, in
which the internal interaction legs could be either Ising type or
Heisenberg interactions, whereas the external legs are necessarily
of the Ising type. This transformation is displayed in figure \ref{fig:3-leg-hybrid-star}.
The limiting case of the transformation illustrated in fig.\ref{fig:3-leg-hybrid-star}
(the Ising coupling) is also known as extended-reduced lattice\cite{Fisher66jmp}.

The Boltzmann factors of the decorated hybrid spin model reads 
\begin{equation}
w(\{s\})=\mathrm{tr}_{\{\sigma\}}\exp\left(H^{XXZ}(\sigma_{1},\sigma_{2},\sigma_{3})+J_{2}(\sigma_{1}^{z}s_{1}+\sigma_{2}^{z}s_{2}+\sigma_{3}^{z}s_{2})\right),
\end{equation}
 where $H^{XXZ}(\sigma_{1},\sigma_{2},\sigma_{3})=\sum_{i=1}^{3}\left\{ J_{1}\left(\sigma_{i}^{x}\sigma_{i+1}^{x}+\sigma_{i}^{y}\sigma_{i+1}^{y}\right)+J_{z}\sigma_{i}^{z}\sigma_{i+1}^{z}\right\} $
with $\sigma_{4}=\sigma_{1}$.

The triangle cell (decorated) could be expressed as Heisenberg coupling,
and as expected, we obtain two configurations for their legs, in correspondence
to the configurations $\uparrow\uparrow\uparrow$ and $\uparrow\uparrow\downarrow$,
so that, in terms of $\varsigma$ we have $\varsigma=3/2$ and $1/2$
respectively. Therefore, we obtain the following Boltzmann factors,
\begin{align}
w(1/2)= & 2\left(\mathrm{e}^{\tfrac{3}{4}J_{z}}+\mathrm{e}^{-\tfrac{1}{4}J_{z}-\tfrac{1}{2}J_{1}}\right)\cosh\left(\frac{J_{2}}{4}\right)+2\mathrm{e}^{\tfrac{1}{4}(-J_{z}+J_{1})}\times,\nonumber \\
 & \left(\mathrm{e}^{-\tfrac{1}{4}J_{2}}\cosh\left(\tfrac{1}{2}\sqrt{J_{2}^{2}+\tfrac{9}{4}J_{1}^{2}+J_{1}J_{2}}\right)+\mathrm{e}^{\tfrac{1}{4}J_{2}}\cosh\left(\tfrac{1}{2}\sqrt{J_{2}^{2}+\tfrac{9}{4}J_{1}^{2}-J_{1}J_{2}}\right)\right)\\
w(3/2)= & 2\mathrm{e}^{\tfrac{3}{4}J_{z}}\cosh\left(\frac{3J_{2}}{4}\right)+2\mathrm{e}^{-\tfrac{1}{4}J_{z}}\left(\mathrm{e}^{J_{1}}+2\mathrm{e}^{-\tfrac{1}{2}J_{1}}\right)\cosh\left(\frac{J_{2}}{4}\right).
\end{align}

\begin{figure}[h]
\subfloat[]{\includegraphics[scale=0.3]{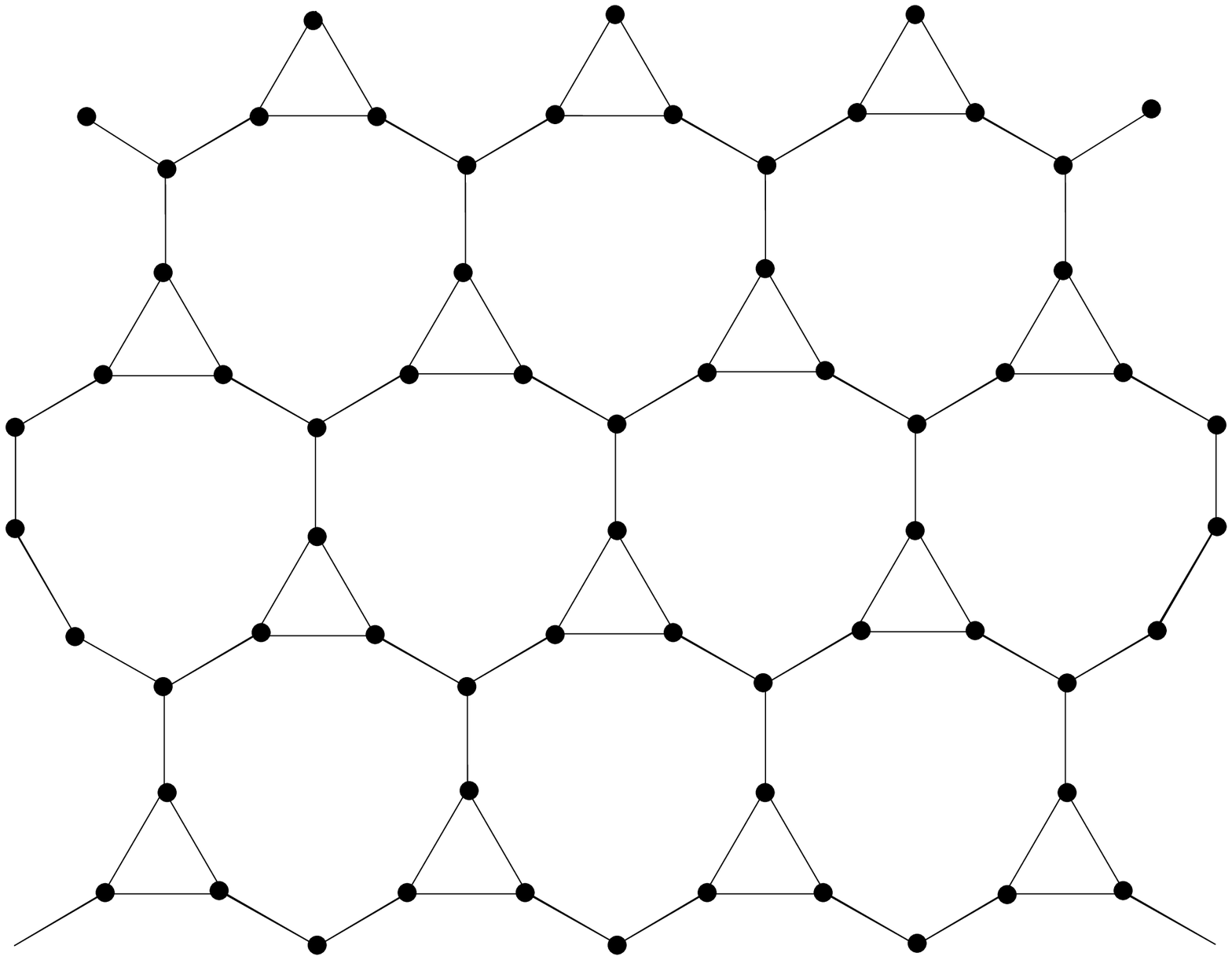}

}\hskip1cm\subfloat[]{\includegraphics[scale=0.35]{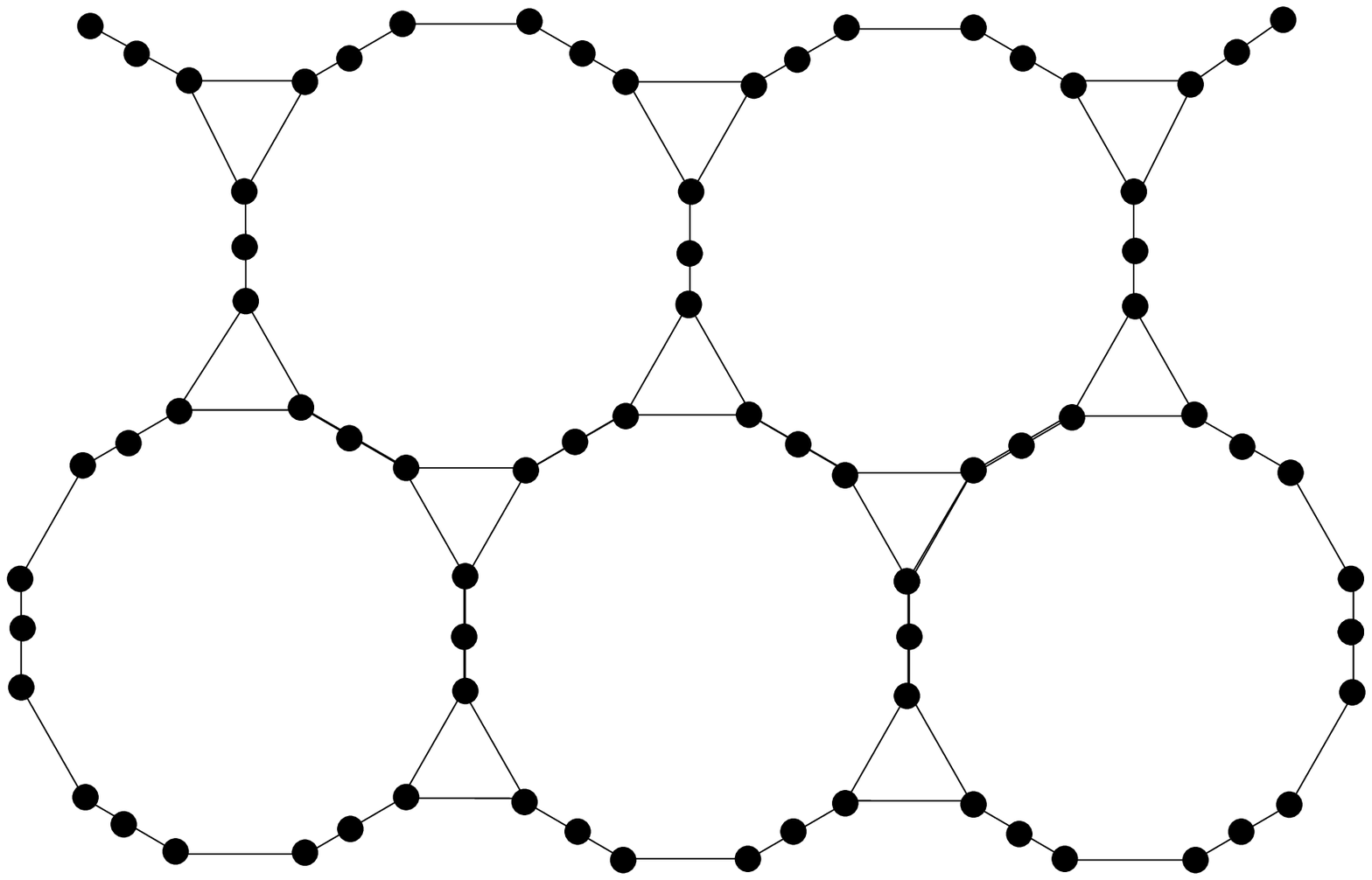}}

\caption{\label{fig:Two-exp-latt}Two possible hybrid lattice, that could be
mapped onto a honeycomb Ising model: (a) the triangle-nonagon (3-9)
Ising-Heisenberg model; and (b) the triangle-dodecagon (3-12) Ising-Heisenberg
model.}
\end{figure}

This hybrid-star transformation could be applied to find the exact
solution of Ising-Heisenberg type models i.e. the 3-9 (triangle-nonagon)
lattice as displayed in figure \ref{fig:Two-exp-latt}(a), where in
its triangle cell we have Heisenberg interaction, whereas in its nonagon
cell we have alternating Ising-Ising-Heisenberg coupling. This lattice
could be mapped onto a honeycomb Ising model\cite{Wu-jmp73,Lin-wu1990}.
In figure \ref{fig:Two-exp-latt}(b) we display the Ising-Heisenberg
model on the 3-12 (triangle-dodecagon) lattice, that can be solved
exactly using the hybrid-star transformation, where once again we
have Heisenberg coupling on triangle cell, and the dodecagon cell
has Ising-Ising-Heisenberg coupling. Detailed discussion about the
thermodynamics properties of these lattice could be analyzed, but
this issue is beyond the scope of this work.

A general expression for hybrid decorated spin model can also be discussed,
where hybrid decoration particle interaction could be expressed in
general by the Hamiltonian $H^{c}(\{\sigma\})$ (such as Heisenberg
interactions), in which $\{\sigma\}$ stands for the set of spin operators
that plays the role of central \textit{mechanical }spin, while the
interaction of central decorated spin with their legs could be given
in general by $H^{l}(\{\sigma\},\{s\}$), in which $\{s\}$ is the
set of Ising spins on the legs. Therefore the general Boltzmann factor
of hybrid spin model could be expressed by
\begin{equation}
w(\{s\})=\mathrm{tr}_{\{\sigma\}}\exp\left(H^{c}(\{\sigma\})+H^{l}(\{\sigma\},\{s\})\right).\label{eq:gen-hyb-H}
\end{equation}

The hybrid-star transformation is equivalent to the hybrid-polygon-star
transformation\cite{PhyscA-09,strecka-pla}; certainly, the parameters
acting on polygons could make the transformation an involving task.

Alternatively the star-star transform can be generalized even when
the legs interact, as we can see in figure \ref{fig:hybd-hybd-trans}(b),where
the transformation not necessarily involves star-like cells. This
transformation could be useful to perform a direct transformation,
such as the square-hexagonal (4-6) lattice\cite{Lin-4-6}, square
octogonal (4-8) lattice\cite{Lin-4-8}, and the pentagonal lattice\cite{urumov}.

\begin{figure}[h]
 \centering{}\includegraphics[scale=0.3]{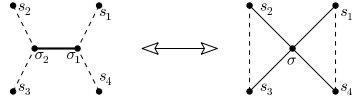}\caption{\label{fig:pent-cell}The alternative hybrid-star like transformation
onto its equivalent uniform spin-1/2 Ising model.}
\end{figure}

The left side Boltzmann's factor of figure \ref{fig:pent-cell} may
be written as 
\begin{equation}
{\displaystyle w(\{s\})={\mathrm{tr}_{\{\sigma\}}}\left\{\exp\left(H^{XXZ}(\sigma_1,\sigma_2)+J[\sigma_{1}(s_{1}+s_{2})+\sigma_{2}(s_{3}+s_{4})]\right)\right\}},\label{eq:alt-pent}
\end{equation}
with $\sigma_1$ and $\sigma_2$ being Pauli matrix.

On the other hand, the Boltzmann factor of transformed plaquette (right
side of figure \ref{fig:pent-cell}) is given by

\begin{equation}
\tilde{w}(\{s\})=\sum_{\sigma=\pm1/2}\exp\left(K{}_{0}+K_{1}\sigma(s_{1}+s_{2}+s_{3}+s_{4})+K_{2}(s_{1}s_{2}+s_{3}s_{4})\right),
\end{equation}
 where $K_{0}$ is a constant shift energy, whereas $K{}_{1}$ is
the interaction parameter between the internal Ising spin $\sigma$ and
each spin $\{s_{i}\}$ and finally $K_{2}$ is the coupling term between
$\{s_{i}\}$, with $\sigma=\pm1/2$ and $s=\pm1/2$.

Imposing the condition $w(\{\tau\})=\tilde{w}(\{\tau\})$, for arbitrary
$\{s_{i}\}$, we obtain only four nonequivalent configurations, namely,
$\{s_{1},s_{2},s_{3},s_{4}\}=$$\{+,+,+,+\}$, $\{+,+,+,-\}$, $\{+,+,-,-\}$
and $\{+,-,+,-\}$. Any other permutation or spin inversion falls
into one of these configurations. Thus the Boltzmann factors reads

\begin{align}
\omega_{1}=w(+,+,+,+)= & 2\mathrm{e}^{K{}_{0}+K{}_{2}/2}\cosh(K_{1}),\label{eq:rot-w1}\\
\omega_{2}=w(+,-,+,-)= & 2\mathrm{e}^{K{}_{0}-K{}_{2}/2},\\
\omega_{3}=w(+,+,-,-)= & 2\mathrm{e}^{K{}_{0}+K{}_{2}/2},\\
\omega_{5}=w(+,+,+,-)= & 2\mathrm{e}^{K{}_{0}}\cosh(K{}_{1}/2).\label{eq:rot-w5}
\end{align}

In order to solve the above equation consistently, the algebraic equation
must satisfy the following relation

\begin{equation}
2\omega_{5}^{2}=(\omega_{1}+\omega_{3})\omega_{2}.\label{Exact-cond}
\end{equation}

After performing some algebraic manipulation of eq. (\ref{eq:rot-w1}-\ref{eq:rot-w5}),
we obtain the magnitudes of the effective interactions this
\begin{align}
\mathrm{e}^{2K{}_{0}}= & \frac{\omega_{3}\omega_{2}}{4},\label{f1}\\
\mathrm{e}^{K{}_{1}}= & \tfrac{\omega_{1}}{\omega_{3}}\pm\sqrt{\left(\tfrac{\omega_{1}}{\omega_{3}}\right)^{2}-1},\\
\mathrm{e}^{K{}_{2}}= & \frac{\omega_{3}}{\omega_{2}}.\label{eq:rel-K2-epsils}
\end{align}

This results is equivalent to that obtained by Urumov\cite{urumov},
using a standard decoration transformation, for more detail the reader
is referred to reference \cite{urumov}, and it can be compared with
our results, showing how the direct transformation avoids the intermediate
transformation. 

It is worth to notice that the relations \eqref{Exact-cond}-\eqref{eq:rel-K2-epsils}
could be valid for any arbitrary spin-$S_{1}$ and spin-$S_{2}$ instead
of $\sigma_{1}$ and $\sigma_{2}$ respectively in eq. \eqref{eq:alt-pent}.
With higher order coupling term on hybrid system, such that satisfy
the spin-inversion symmetry.

\subsection{Hybrid-hybrid transformation}

\begin{figure}[h]
\subfloat[]{\includegraphics[scale=0.3]{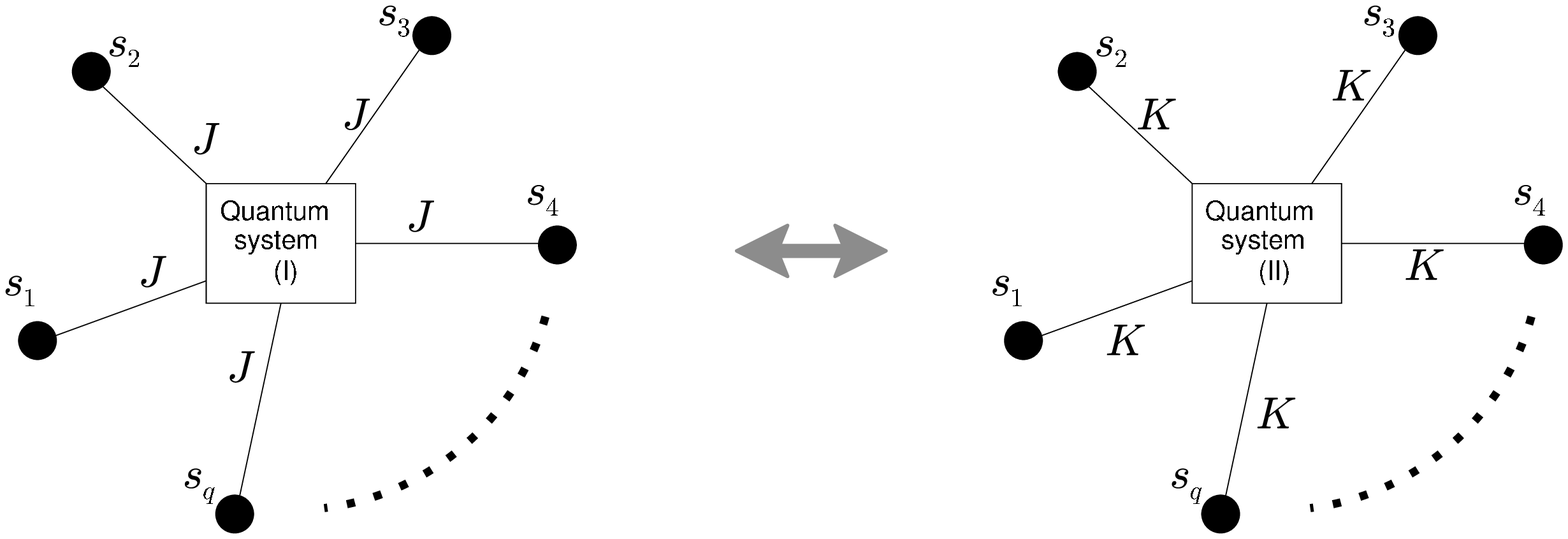}

}\hskip1cm\subfloat[]{\includegraphics[scale=0.3]{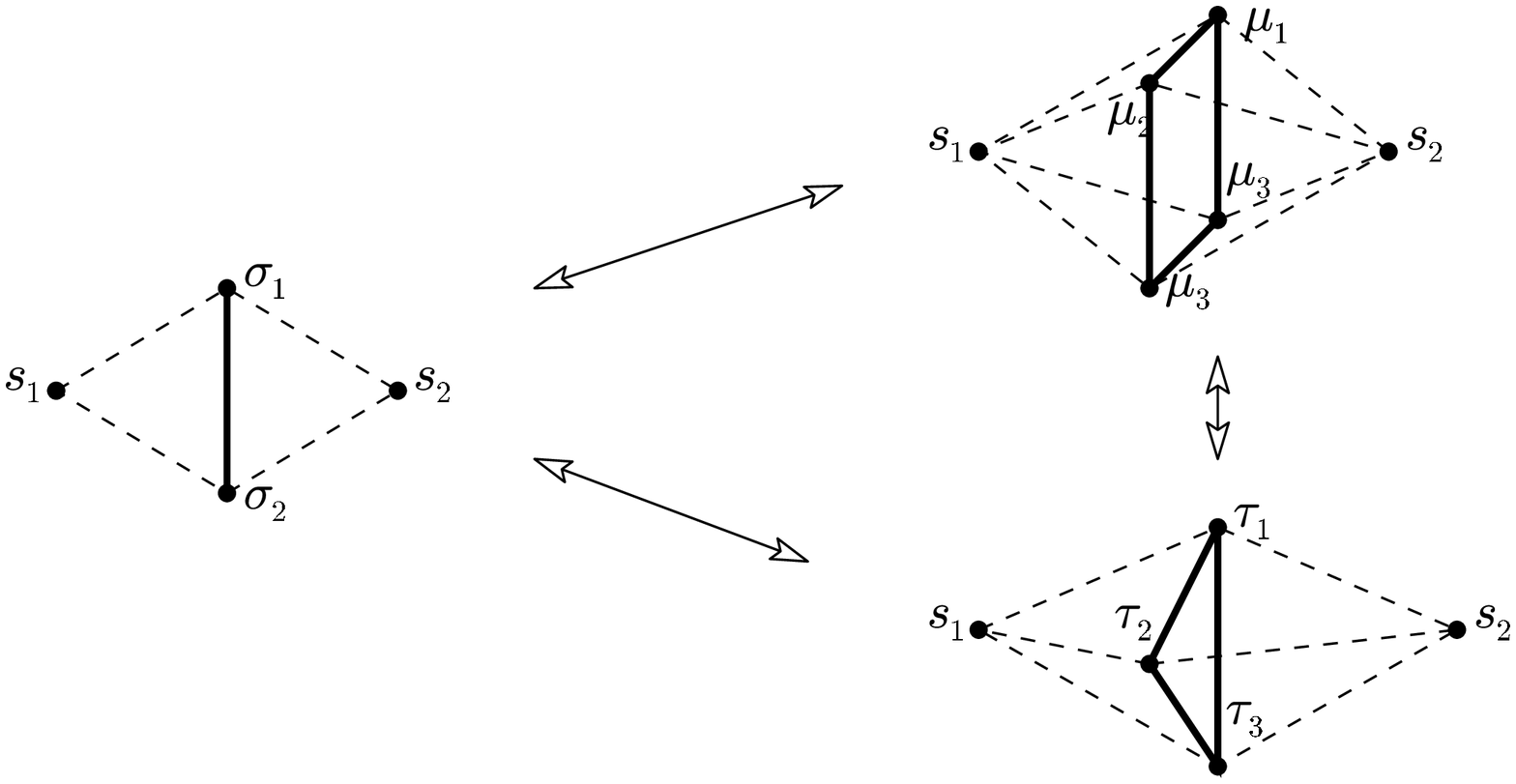}

}

\caption{\label{fig:hybd-hybd-trans}(a) The hybrid-hybrid decorated transformation,
onto its equivalent hybrid system. (b) Example of hybrid-hybrid decoration
transformation.}
\end{figure}

In general it is still possible to extend the decoration transformation
to hybrid-hybrid transformation. This kind of transformation could
be for the direct mapping of some hybrid model, such as given by the
Hamiltonian \eqref{eq:gen-hyb-H} onto another hybrid model with different
topological structure. Physically, this could help the understanding
of physical properties of two different hybrid models (see fig. \ref{fig:hybd-hybd-trans}(a)).
Thus the Boltzmann factor of the effective hybrid spin model may be
expressed by 
\begin{equation}
\tilde{w}(\{s\})=\mathrm{tr}_{\{\tau\}}\exp\left(K_{0}+\tilde{H}^{c}(\{\tau\})+\tilde{H}^{l}(\{\tau\},\{s\})\right),
\end{equation}
 where $\tilde{H}^{c}(\{\tau\})$ is the Hamiltonian of the central
\textit{mechanical} spin, $\{\tau\}$ are the spin operators of effective
lattice that interacts inside the \textit{mechanical} spin, while
the Hamiltonian $\tilde{H}^{l}(\{\tau\},\{s\})$ represents the interaction
of the central \textit{mechanical} spin and its legs with spins $\{s\}$.

As an illustrative example that we consider, could be the displayed
in figure \ref{fig:hybd-hybd-trans} (b), the Boltzmann factor for
the decorated spin model becomes

\begin{equation}
w(\{s\})=\mathrm{tr}_{\{\sigma\}}\left(\mathrm{e}^{H^{XXZ}(\sigma_{1},\sigma_{2})+J_{2}(\sigma_{1}^{z}+\sigma_{2}^{z})(s_{1}+s_{2})}\right),\label{eq:w-hybrd-dec}
\end{equation}
 whereas the effective Boltzmann factor will be given by eq. \eqref{eq:w-hybrd-eff}

\begin{equation}
\tilde{w}(\varsigma)=2\mathrm{e}^{K_{z}/4}\cosh(K_{2}\varsigma)+2\mathrm{e}^{-K_{z}/4}\cosh(K_{1}/2).\label{eq:w-hybrd-hrbrd}
\end{equation}

Imposing that eq.\eqref{eq:w-hybrd-dec} and \eqref{eq:w-hybrd-hrbrd}
be equivalent we obtain the following relations,

\begin{equation}
\mathrm{e}^{K_{z}/4}=\frac{w(1)-w(0)}{2\left(\cosh(K_{2})-1\right)},
\end{equation}

\begin{equation}
K_{1}=2\text{arcosh}\left(\mathrm{e}^{K_{z}/4}\left(\tfrac{w(0)}{2}-\mathrm{e}^{K_{z}/4}\right)\right).
\end{equation}

Note that the parameter $K_{2}$ is an independent parameter in this
transformation.

The hybrid-hybrid transformation is equivalent to the standard hybrid-polygon-hybrid
decoration transformation \cite{PhyscA-09,strecka-pla}; clearly,
the algebraic manipulations involved in the hybrid-hybrid transformation
are much easier to perform.

\section{Conclusions}

In this paper we present a direct transformation for a general decorated
spin model. We have discussed that the advantage of this transformation
is that of avoiding the proliferation of unnecessary intermediate
parameter which only makes the algebraic calculation cumbersome. We
have discussed the $q$-leg star-star transformation with any central
\textit{mechanical} spin and spin-1/2 particles on their legs, thus
finding that the transformation will be possible for $q\geqslant4$,
only if the decorated spin model satisfy the conditions given in eqs.
\eqref{eq:ws-cond-even-e}-\eqref{eq:ws-cond-odd} and spin-inversion
symmetry. When spin-inversion symmetry is not satisfied, this conditions
becomes a more involving relation. The case of higher order spins
has been also discussed, and we show that the expression of parameter
constraints becomes more cumbersome. Finally the extension of decoration
transformation to classical-quantum (hybrid) spin models has been
discussed as well, in which several decorated hybrid spin models could
be mapped onto other hybrid spin models with different topology. All
transformation discussed above are illustrated by several examples
and most of these models were not explored yet, therefore it could
be interesting to discuss this models in order to study its thermodynamic
properties.

\section*{Acknowledgments}

The authors thank E.V. Corrêa Silva, for his careful reading of our
manuscript. This work was supported partially by CNPq and FAPEMIG.

\end{document}